\documentclass[prd,twocolumn,floatfix,amsmath,amssymb,floatfix,nofootinbib]{revtex4}
\usepackage{graphicx,color,dcolumn,booktabs,bm}
\usepackage{longtable,lscape,comment,braket}
\usepackage{txfonts}
\usepackage{overpic}
\usepackage{amssymb}
\usepackage{indentfirst}
\usepackage{feynmf}   %{feynmp}
\usepackage{slashed}  %for Feynman symbols
\usepackage{cases}
\usepackage{epstopdf}
\usepackage{psfrag}
\usepackage{subfigure}
\usepackage{threeparttable}
\usepackage{color}
\usepackage{gensymb}
\usepackage{multirow}
\usepackage{graphicx,color,dcolumn,booktabs,bm}
\usepackage[title]{appendix}
\graphicspath{{Figures/}} %
\usepackage[colorlinks, citecolor=blue,anchorcolor=red,menucolor=red, linkcolor=red,filecolor=red,runcolor=red,urlcolor=blue,frenchlinks=red]{hyperref}

\usepackage{ulem}

\makeatletter
\renewcommand{\maketag@@@}[1]{\hbox{\m@th\normalsize\normalfont#1}}%
\makeatother
\begin{document}
%\begin{CJK}{GBK}{}

\title{The role of triangle singularity in isospin breaking process $J/\psi \to \Lambda \bar{\Lambda} \pi$ and the possible evidence of $\Sigma^*(\frac{1}{2}^-)$ states}
\author{Qi Huang$^{1}$}\email{huangqi@nnu.edu.cn}
\author{Zi-Xuan Ma$^{1}$}\email{19190101@njnu.edu.cn}
\author{Jia-Jun Wu$^{2,3}$}\email{wujiajun@ucas.ac.cn}\thanks{Corresponding author}
\author{Rong-Gang Ping$^{2,4}$}\email{pingrg@ihep.ac.cn}\thanks{Corresponding author}
\author{Jun He$^{1}$}\email{junhe@njnu.edu.cn}
\author{Hong-Xia Huang$^{1}$}\email{hxhuang@njnu.edu.cn}
\affiliation{$^1$School of Physics and Technology, Nanjing Normal University, Nanjing 210023, China\\
$^2$University of Chinese Academy of Sciences (UCAS), Beijing 100049, China\\
$^3$Southern Center for Nuclear-Science Theory (SCNT), Institute of Modern Physics, Chinese Academy of Sciences, Huizhou 516000, China\\
$^4$Institute of High Energy Physics, Chinese Academy of Sciences, P.O. Box 918(1), Beijing 100049, China}

\begin{abstract}

In this study, the impact of triangle singularity is investigated in the isospin-breaking process $J/\psi \to \Lambda \bar{\Lambda} \pi$.
The triangle singularity is found to play a significant role in the process, resulting in the creation of a resonance-like structure around 1.4 GeV in the $\Lambda\pi(\bar{\Lambda}\pi)$ invariant mass spectrum.
To amplify the impact of this triangle singularity, the presence of two $\Sigma^*(\frac{1}{2}^-)$ states around 1.4 GeV and 1.6 GeV is essential, yet these states have not been definitively identified in the current baryon spectrum. 
We recommend that experiments, particularly the Beijing Spectrometer (BESIII) and the future Super Tau-Charm Factory (STCF), to investigate the process $J/\psi \to \Lambda \bar{\Lambda} \pi$ to offer direct evidences for our predicted triangle singularity and additional evidence regarding the $\Sigma(\frac{1}{2}^-)$ states.

\end{abstract}

\maketitle

\section{Introduction}\label{sec1}

The current BESIII experiment has accumulated a wealth of data, and future STCF experiments are expected to have higher luminosity. 
This makes the experiment in the tau-charm energy region a crucial location for studying non-perturbative QCD through precise measurements of specific QCD processes~\cite{BESIII:2020nme, Achasov:2023gey}.
Even more intricate structures could be unveiled if a sufficient number of events are available in the future.
For example, the existence of a triangle singularity is one of the most fascinating topics~\cite{Shen:2020gpw}, representing a unique feature of final state interaction when a particular kinematic configuration, referred to as the Coleman-Norton theorem.
Although proposed by L. D. Landau in 1959~\cite{Landau:1959fi}, the concept of triangle singularity may hold significant importance in elucidating various anomalies~\cite{Shen:2020gpw,Wu:2011yx, Aceti:2012dj, Wu:2012pg, Ketzer:2015tqa, Wang:2013cya,Wang:2013hga, Achasov:2015uua, Liu:2015taa,Liu:2015fea,Guo:2015umn,Szczepaniak:2015eza, Guo:2016bkl, Bayar:2016ftu, Wang:2016dtb, Pilloni:2016obd, Xie:2016lvs, Szczepaniak:2015hya, Roca:2017bvy,
Debastiani:2017dlz, Samart:2017scf, Sakai:2017hpg, Pavao:2017kcr, Xie:2017mbe, Bayar:2017svj,Liang:2017ijf, Oset:2018zgc, Dai:2018hqb, Dai:2018rra, Guo:2019qcn, Liang:2019jtr, Nakamura:2019emd,Liu:2019dqc, Jing:2019cbw, Braaten:2019gfj, Sakai:2020ucu, Sakai:2020fjh, Molina:2020kyu, Braaten:2020iye, Alexeev:2020lvq, Ortega:2020ayw,Du:2019idk,Liu:2020orv,Achasov:2019wvw,Huang:2020kxf,Huang:2021olv,Luo:2021hyy,Wang:2022wdm,Lu:2022kdh,He:2023plt,Wang:2023xua} (a comprehensive review can be found in Ref.~\cite{Guo:2019twa}). 
However, as highlighted by Refs.~\cite{Huang:2020kxf,Huang:2021olv,COMPASS:2020yhb}, the concept of triangle singularity itself has not been definitively examined.

Based on previous studies, three main factors make it challenging to examine triangle singularities in experiments: interference with threshold cusps, mixing with resonances, and the lack of precise predictions due to unknown vertices~\cite{Liu:2015taa,Huang:2020kxf}. 
Therefore, in response to these challenges and with a focus on BESIII and STCF, Refs.~\cite{Huang:2020kxf,Huang:2021olv,Wang:2022wdm,Wang:2023xua}  have proposed several processes in which the predicted pure effects of triangle singularities could potentially be examined in the future.

Upon reviewing previous works~\cite{Huang:2020kxf, Huang:2021olv}, we have identified two inherent weaknesses, stemming from the discussions on the background generated by the tree diagrams. 
One issue is the potential omission of several tree diagrams, leading to inaccuracies in estimating the background.
In this scenario, the significance of the triangle singularity is not entirely clear. 
The other concern is that the signal of the triangle singularity is significantly smaller than the contribution from the background channel. 
This implies that the signal of the triangle singularity can only have an observable effect when it interferes with the background channel. 
Therefore, in previous calculations, it was always essential to treat an interference angle between the background and TS signal as a free parameter, making it impossible to determine accurately. 
Therefore, as emphasized in our previous works~\cite{Huang:2020kxf,Huang:2021olv}, we assumed the phase angle to be zero.
This implies that for certain specific phase angles, the effect of the triangle singularity may be significantly suppressed, making it still unobservable in experiments.

Therefore, we need to develop improved processes to assist experiments in detecting the signals of triangle singularities. 
Inspired by the concept introduced in previous studies~\cite{Wu:2011yx, Aceti:2012dj, Wu:2012pg, Achasov:2015uua}, where the triangle singularity effect plays a crucial role in the isospin-breaking process $\eta(1405/1475) \to \pi f_0(980) \to 3\pi$, this study focuses on the $J/\psi \to \Lambda \bar{\Lambda} \pi$ process. 
The inner triangle loop in this process involves $\Sigma^\ast~(\bar{\Sigma}^\ast)$, $\bar{\Sigma}~(\Sigma)$, and $\pi$, and we have chosen this specific process for the following reasons:

\textbf{Within the triangle loop, the triangle singularity and isospin-breaking effect occur simultaneously.}
Given that the isospins of $J/\psi$, $\Lambda~(\bar{\Lambda})$, and $\pi$ are 0, 0, and 1, respectively, the isospin is not conserved in this process. 
Clearly, if this process occurs through strong interactions obeying the isospin conservation, the outcome should be zero.
Nevertheless, due to the mass discrepancies among $\Sigma^{(\ast)}$ particles with varying charges~\cite{ParticleDataGroup:2022pth}, the $\Sigma^{(\ast)} \bar{\Sigma} \pi$ loops with different charge states will not cancel each other out. 
More significantly, the triangle singularities will also occur in these loops.
Consequently, the locations of the triangle singularities will be distinct from each other as a result of the mass variations of $\Sigma^{(\ast)}$, as illustrated in Fig.~\ref{fig:difference} below. 
Therefore, this isospin-breaking process may exhibit a resonance-like structure around 1.4 GeV in the $\Lambda\pi~(\bar{\Lambda}\pi)$ invariant mass spectrum, as shown in Fig.~\ref{fig:difference}.
This scenario bears a striking resemblance to the $J/\psi \to \gamma\eta(1405/1475)\to \gamma\pi\pi\pi$ process via the $K^*\bar{K}K$ triangle loop.

\textbf{Reduction of Background from Isospin Breaking.} 
Given that the $J/\psi \to \Lambda \bar{\Lambda} \pi$ process involves isospin breaking, the tree diagrams associated with this process are expected to be suppressed. 
Consequently, the background of this process will be significantly reduced compared to conventional scenarios.
This reduction in background enhances the prominence of the triangle singularity effect. 
As a result, the interference effects may be less significant, and the phase angle is less likely to introduce uncertainty when observing the triangle singularity signal.

\textbf{A large number of $J/\psi$ events have been accumulated in BESIII and are anticipated in future STCF experiments.} 
In comparison to the $\psi(2S)$ selection used in our previous studies~\cite{Huang:2020kxf, Huang:2021olv}, the $J/\psi$ is the most prevalent charmonium worldwide.
As of now, BESIII has accumulated over $10^9$ $J/\psi$ events~\cite{BESIII:2020nme}, and in the future STCF, the number of $J/\psi$ events is expected to be even greater, given that the luminosity of STCF is approximately $10^2$ times higher than that of BESIII~\cite{Achasov:2023gey}.
Therefore, the present and upcoming experimental settings offer an excellent platform for conducting more accurate measurements in $J/\psi$ physics. 
This suggests that the observation of the triangle singularity in the $J/\psi \to \Lambda \bar{\Lambda} \pi$ process is more likely to occur in the future.

Moreover, the properties of the resonances participating in the triangle loop are also crucial in augmenting the significance of the triangle singularity.
For instance, when the triangle singularity occurs, the relative momentum between particles is not very high, leading to a significant suppression of high partial wave interactions.
So the triangle singularity becomes a valuable probe for testing the existence of a resonance that can decay into the final state in an $S$-wave configuration.
Thus, in the $J/\psi \to \Lambda \bar{\Lambda} \pi$ process, it provides a nice place to examine the existence of the $\Sigma^{\ast}$ particles with masses around 1.4 GeV and $J^P=\frac{1}{2}^-$, which decays into $\Lambda\pi$ in $S-$wave.
Fortunately, based on previous studies, there is a potential candidate, the $\Sigma(1381)$~\cite{Zhang:2004xt, Wu:2009tu, Wu:2009nw}, suggested by a multiquark model, which could play a significant role in this process. 
The clear contribution of the triangle singularity in this process could support its existence.

The remainder of this paper is structured as follows: 
In Section \ref{sec2}, a brief introduction to the model will be provided. 
Section \ref{sec3} will present the numerical results and the corresponding discussions.
Finally, a summary will be provided in Section \ref{sec4}.

\section{Model setup}\label{sec2}
\begin{figure}[hhtbp]
  \centering
  \includegraphics[width=1.0\linewidth]{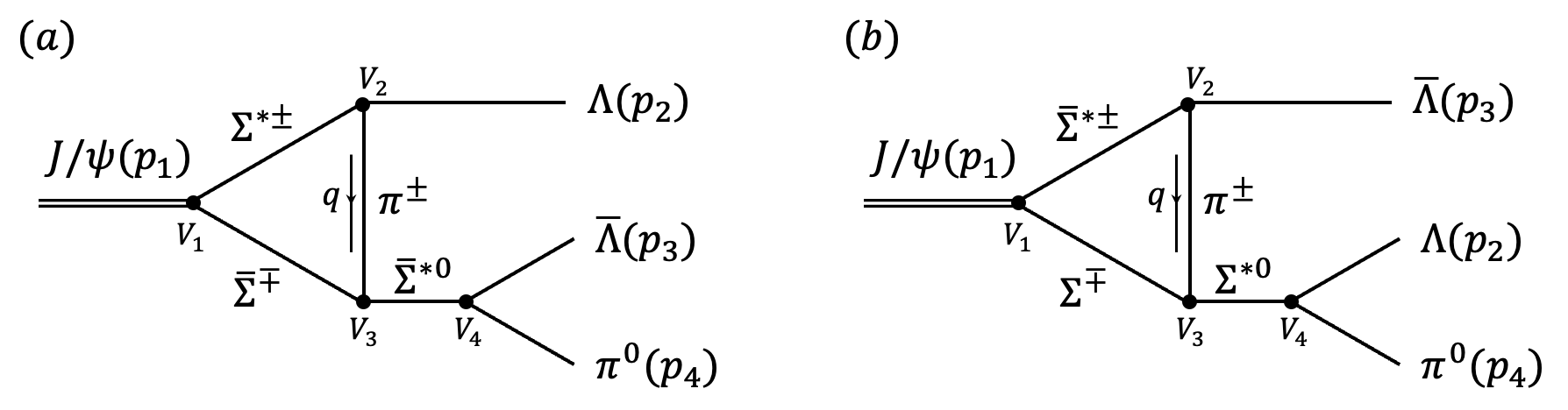}
\caption{The decay mechanisms of $J/\psi \to \Lambda \bar{\Lambda} \pi^0$ process via triangle loops, where $(a)$ and $(b)$ are just conjugate to each other.}
\label{fig:mechanism}
\end{figure}

The mechanisms involving the one triangle loop of the $J/\psi \to \Lambda \bar{\Lambda} \pi$ process are illustrated in Fig.~\ref{fig:mechanism}, where Fig.~\ref{fig:mechanism} (a) and Fig.~\ref{fig:mechanism} (b) are actually conjugate to each other. 
In this mechanism, the $J/\psi$ decays initially into $\Sigma^{(\ast)}~(\bar{\Sigma}^\ast)$ and $\bar{\Sigma}~(\Sigma)$, followed by the decay of $\Sigma^{(\ast)}~(\bar{\Sigma}^\ast)$ into $\Lambda\pi~(\bar{\Lambda}\pi)$, while the $\bar{\Sigma}~(\Sigma)$ remains unchanged. 
Finally, $\pi$ catches $\bar{\Sigma}~(\Sigma)$ and re-scatters into $\bar{\Lambda}\pi~(\Lambda\pi)$ via an excited $\bar{\Sigma}^\ast~(\Sigma^\ast)$. 
Therefore, when a triangle singularity occurs, all the particles forming the triangle loop are on their mass shells, resulting in a peak in the $\Lambda\pi~(\bar{\Lambda}\pi)$ invariant mass spectrum.

Similar to our previous studies~\cite{Huang:2020kxf, Huang:2021olv}, we utilize the effective Lagrangian method to conduct the entire calculation. 
This approach allows us to express the general form of the amplitude that characterizes the loop diagram in Fig.~\ref{fig:mechanism} as:
\begin{eqnarray}
  \mathcal{M}^{\mathrm{Loop}} &=& i\int \frac{d^4 q}{(2\pi)^4}\frac{V_3 \mathcal{F}(p_3+p_4-q, m_\Sigma, \Lambda_\Sigma)} {(p_3+p_4-q)^2-m_\Sigma^2+im_\Sigma\Gamma_\Sigma}\nonumber\\
  &&\times \frac{V_1 \mathcal{F}(p_2+q,m_{\Sigma^\ast},\Lambda_{\Sigma^\ast})} {(p_2+q)^2-m_{\Sigma^\ast}^2+im_{\Sigma^\ast}\Gamma_{\Sigma^\ast}} \nonumber \\
  &&\times\frac{V_2 \mathcal{F} (q,m_\pi,\Lambda_\pi)} {q^2-m_\pi^2+i m_\pi \Gamma_\pi} \nonumber\\ 
  &&\times \frac{V_4 \mathcal{F}(p_3+p_4,m_{\Sigma^\ast},\Lambda_{\Sigma^\ast})}{(p_3+p_4)^2-m_{\Sigma^\ast}^2+im_{\Sigma^\ast}\Gamma_{\Sigma^\ast}}, \label{eq:loop}
\end{eqnarray}
where $V_i$ are the interactions of each vertices as given in Fig.~\ref{fig:mechanism}. 
For the interactions between $J/\psi$ and $\Sigma^{(\ast)} \bar{\Sigma}^{(\ast)}$, the relevant effective Lagrangians are~\cite{Tsushima:1996xc, Tsushima:1998jz, Zou:2002yy, Ouyang:2009kv, Wu:2009md, Cao:2010km, Cao:2010ji}
\begin{eqnarray}
  \mathcal{L}_{\psi P_{11} \bar{P}_{11}} &=&-g_{\psi P_{11} \bar{P}_{11}} \bar{P}_{11} \gamma_\mu \psi^\mu P_{11} + h.c.,\label{eq:Lagrangian-P11-V}\\
  \mathcal{L}_{\psi S_{11} \bar{P}_{11}} &=& -g_{\psi S_{11} \bar{P}_{11}} \bar{P}_{11} \gamma_5 \gamma_\mu \psi^\mu S_{11} + h.c.,\label{eq:Lagrangian-S11-V}\\
  \mathcal{L}_{\psi D_{13} \bar{P}_{11}} &=& -g_{\psi D_{13} \bar{P}_{11}} \bar{P}_{11} \psi_\mu D_{13}^\mu + h.c.,\label{eq:Lagrangian-D13-V}\\
  \mathcal{L}_{\psi P_{13} \bar{P}_{11}} &=& -i g_{\psi P_{13}\bar{P}_{11}} \bar{P}_{11} \gamma^5 \gamma^\nu \left( \partial_\mu \psi_\nu - \partial_\nu \psi_\mu \right) P_{13}^\mu + h.c.,\nonumber\\
  \label{eq:Lagrangian-P13-V}
\end{eqnarray}
with $P_{11}$, $S_{11}$, $D_{13}$, $P_{13}$ being the fields of excited baryons with quantum numbers $J^P = 1/2^+$, $1/2^-$, $3/2^-$, and $3/2^+$ respectively. 
For the interactions between baryons and pion, the effective Lagrangians are~\cite{Xu:2015qqa}
\begin{eqnarray}
  \mathcal{L}_{\pi P_{11} P_{11}}&=&-g_{\pi P_{11} P_{11}} \bar{P}_{11} \gamma_5 \gamma_\mu \tau\cdot\partial^\mu\pi P_{11} + h.c.,\label{eq:Lagrangian-P11-P}\\
  \mathcal{L}_{\pi P_{11} S_{11}}&=&-g_{\pi P_{11} S_{11}} \bar{P}_{11} \tau\cdot\pi S_{11} + h.c.,\label{eq:Lagrangian-S11-P}\\
  \mathcal{L}_{\pi P_{11} D_{13}}&=&-g_{\pi P_{11} D_{13}} \bar{P}_{11} \gamma_5 \gamma^\mu \tau\cdot\partial_\mu\partial_\nu\pi D_{13}^\nu + h.c.\label{eq:Lagrangian-D13-P},\\
  \mathcal{L}_{\pi P_{11} P_{13}} &=& g_{\pi P_{11} P_{13}} \bar{P}_{11} \tau\cdot\partial_\mu\pi P_{13}^\mu + h.c..\label{eq:Lagrangian-P13-P}
\end{eqnarray}

In Eq. (\ref{eq:loop}), a form factor ${\mathcal{F}(q,m,\Lambda) = \frac{\Lambda^4}{(q^2-m^2)^2+\Lambda^4}}$ is introduced to describe the structure effects of interaction vertices and off-shell effects of internal particles. 
Furthermore, it can also avoid the ultraviolet divergence appeared in our calculation. 
As emphasized in our previous work, when triangle singularity happens, all the particles that composing the triangle loop are on-shell, leading to 
${\mathcal{F}(q,m,\Lambda) = 1}$.
This indicate that the form factor will not affect the height of the peak caused by the triangle singularity very much~\cite{Huang:2020kxf,Huang:2021olv}. 
Furthermore, the form factor ${\mathcal{F}(q,m,\Lambda)}$ includes a free parameter $\Lambda$, representing the momentum cutoff and commonly parameterized as $\Lambda=m+\alpha \Lambda_{\text{QCD}}$. 
In our following calculation, we take $\Lambda_{\text{QCD}} = 0.22$ GeV.
Since $\alpha$ affects little on the behavior of $\mathcal{M}^{\mathrm{Loop}}$, it will still be taken as 1~\cite{Huang:2020kxf}.

The Dalitz plots and invariant mass spectra we need can be obtained by
\begin{eqnarray}
  d\Gamma = \sum \frac{\left|\mathcal{M}^{\mathrm{Loop}}\right|^2}{96(2\pi)^3 m_{J/\psi}^3} dm_{23}^2 dm_{34}^2,
\end{eqnarray}
or
\begin{eqnarray}
  d\Gamma = \sum \frac{\left|\mathcal{M}^{\mathrm{Loop}}\right|^2}{96(2\pi)^3 m_{J/\psi}^3} dm_{24}^2 dm_{34}^2,
\end{eqnarray}
where $\sum$ denotes the summations over spins of initial and final particles, and the invariant masses are defined as $m^2_{23}=(p_2+p_3)^2$, $m^2_{24}=(p_2+p_4)^2$, $m^2_{34}=(p_3+p_4)^2$~\cite{ParticleDataGroup:2022pth}.

\section{Numerical results and discussions}\label{sec3}

\subsection{Triangle singularity effect caused by difference charge configurations in the isospin breaking process $J/\psi \to \Lambda \bar{\Lambda} \pi$}

\begin{figure}[htbp]
  \centering
  \includegraphics[width=1.0\linewidth]{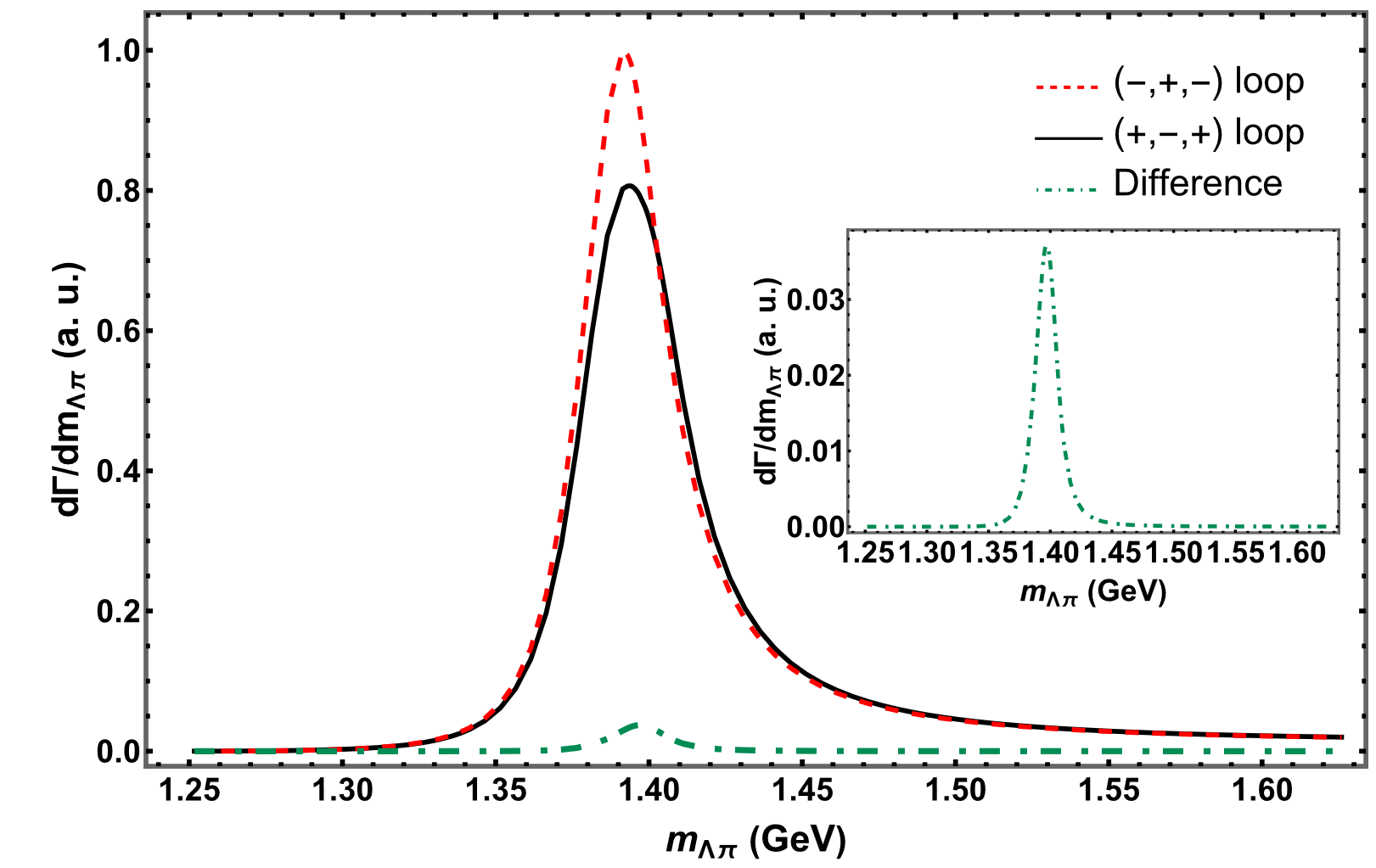}
\caption{The $m_{\Lambda\pi}$ invariant mass spectra of the $J/\psi \to \Lambda \bar{\Lambda} \pi$ through Fig.~\ref{fig:mechanism} $(b)$, where $(\pm,\mp,\pm)$, i.e. black solid line and red dashed line, denotes that the triangle loop is composed by $\bar{\Sigma}^{\ast\pm} \Sigma^\mp \pi^\pm$. 
The enlarged subfigure with green dot-dashed line denotes the total contribution of these two kinds of charge configurations, which represents the difference between them. 
It should be noticed here that in this figure the maximum of the value has been normalized to 1.0, and $\bar{\Sigma}^{\ast\pm} = \bar{\Sigma}^\ast (1670)^\pm$, $\Sigma^{\ast 0} = \Sigma^\ast (1381)^0$.}
\label{fig:difference}
\end{figure}

With the preparations given in Sec.~\ref{sec3}, we proceed to present the numerical calculations and corresponding discussions. 
In Fig.~\ref{fig:mechanism}, the charge configuration with primary decay process where $J/\psi$ decaying into two neutral $\Sigma^{(\ast)0}$ particles is not included. 
It is because under this situation the isospin factor of $V_3$ vertex, i.e., the interaction between $\Sigma^{\ast0}$ and $\Sigma^0 \pi^0$, is zero, due to the Clebsch-Gorden coefficient of isospin factor $\langle 1,0;1,0|1,0 \rangle = 0$.
Thus, the neutral channel will have no contribution to the process. 
For a single loop diagram in Fig.~\ref{fig:mechanism}, after multiplying the isospin factors of all the vertices $V_i (i=1, ... ,4)$, the signs of the amplitudes of the two kinds of charge configurations in the triangle loop are opposite. 
Therefore, if the $\Sigma^{(\ast)}$ particles have no mass differences in charges, the contributions of these two charge configurations will cancel each other and make the total amplitude of the process still be zero. 
However, if we take into account the practical masses differences of charge states~\cite{ParticleDataGroup:2022pth}, we will find some changes on the lineshapes in $\Lambda\pi~(\bar{\Lambda}\pi)$ invariant mass spectrum, as shown in Fig.~\ref{fig:difference}.

In Fig.~\ref{fig:difference}, for instance, we choose Fig.~\ref{fig:mechanism} $(b)$ to illustrate the lineshapes of the $m_{\Lambda\pi}$ invariant mass spectrum. 
Here, $\bar{\Sigma}^{\ast\pm}$ and $\Sigma^{\ast 0}$ are assigned as $\bar{\Sigma}^\ast (1670)^\pm$ and $\Sigma^\ast (1381)^0$, respectively, with all coupling constants set to 1. 
In RPP~\cite{ParticleDataGroup:2022pth}, the masses of $\Sigma^-$ and $\Sigma^+$ are 1197.449 MeV and 1189.37 MeV, respectively, and the mass of $\Sigma^\ast(1670)$ is just around 1662 MeV~\cite{ParticleDataGroup:2022pth}. 
After applying Coleman-Norton theorem %\cite{Shen:2020gpw,Huang:2020kxf}
, the positions of triangle singularities are given as 1.40909 GeV and 1.40073 GeV for the different kinds of charge intermediate states, respectively. 
From Fig.~\ref{fig:difference}, it can be observed that the lineshapes of the black solid line and the red dashed line exhibit slight differences in the following two aspects.
One difference is that the central value of the peak in the black solid line is slightly larger than that in the red dashed line due to variations in the positions of the triangle singularities.
Another distinction is that the height of the black solid line is slightly lower than that of the red dashed line, attributed to differences in the phase spaces.
Crucially, the non-zero total contribution of the green dot-dashed line arises from the mass distinction of $\Sigma^\mp$ in these two charge configurations. 
Consequently, a resonance-like structure emerges around 1.4 GeV in the $\Lambda\pi$ invariant mass spectrum, with a width of approximately 20 MeV.

\subsection{Discussions on the Dalitz plots with different $\Sigma^{\ast0}$ selections}

\begin{figure}[htbp]
  \centering
  \includegraphics[width=1.0\linewidth]{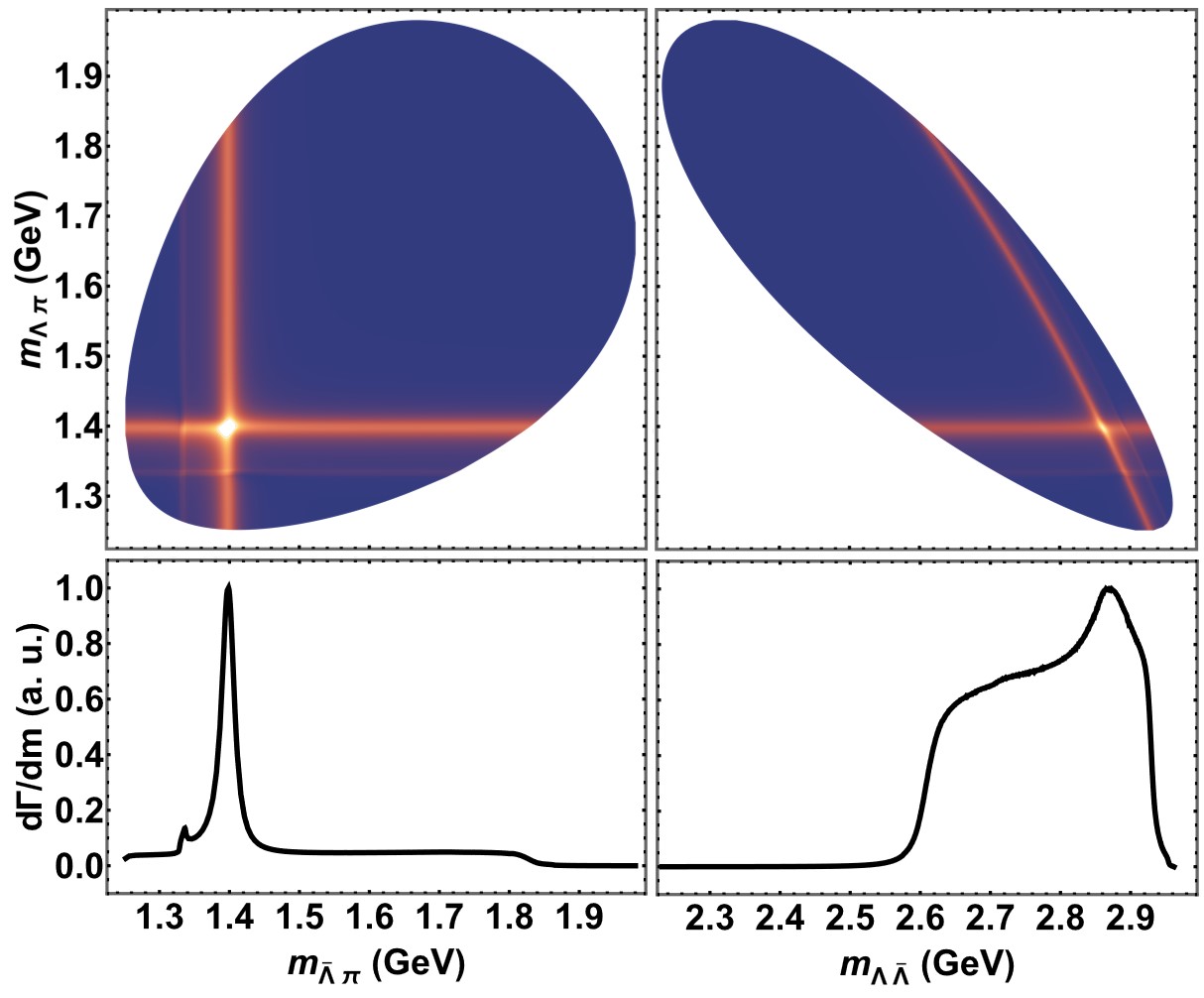}
\caption{
The Dalitz plots in addition with their projections on $\Lambda \pi$ and $\Lambda \bar{\Lambda}$ invariant masses of $J/\psi \to \Lambda \bar{\Lambda} \pi$ process, where both the two terms presented in Fig.~\ref{fig:mechanism} are considered.
The $\Sigma^{\ast \pm}~(\bar{\Sigma}^{\ast \pm})$ is set as $\Sigma^\ast(1670)^\pm~(\bar{\Sigma}^\ast(1670)^\pm)$, while the $\Sigma^{\ast 0}~(\bar{\Sigma}^{\ast 0})$ is set as $\Sigma^\ast(1381)^0~(\bar{\Sigma}^\ast(1381)^0)$. 
For the four subfigures, the top-left is Dalitz plot of $\bar{\Lambda} \pi - \Lambda \pi$ invariant mass spectra, and the top-right is Dalitz plot of $\Lambda \bar{\Lambda} - \Lambda \pi$ invariant mass spectra, and the bottom-left is distribution of $d\Gamma/dm_{\bar{\Lambda}\pi}$, and the bottom-right is distribution of $d\Gamma/dm_{\Lambda \bar{\Lambda}}$. 
It should be noticed here that both the maximum values of the distributions in the bottom row have been normalized to 1.0.}
\label{fig:dalitz16701381}
\end{figure}

In Fig.~\ref{fig:difference}, we only take into account the contribution of Fig.~\ref{fig:mechanism} $(b)$. 
Given that Fig.~\ref{fig:mechanism} $(a)$ is simply the conjugate of Fig.~\ref{fig:mechanism} $(b)$, it is conceivable that the lineshapes of the $\bar{\Lambda}\pi$ invariant mass spectrum will be identical if we solely consider the contribution from Fig.~\ref{fig:mechanism} $(a)$. 
To account into all the contributions, we should sum the amplitudes of these two diagrams coherently, and the interference effects between them are necessary. 
In Fig.~\ref{fig:dalitz16701381}, we give the Dalitz plots in addition with their projections on the invariant masses, where both the two terms in Fig.~\ref{fig:mechanism} are included.
In Fig.~\ref{fig:mechanism}, the $\Sigma^{\ast \pm}~(\bar{\Sigma}^{\ast \pm})$ is set as $\Sigma^\ast(1670)^\pm~(\bar{\Sigma}^\ast(1670)^\pm)$, and the $\Sigma^{\ast 0}~(\bar{\Sigma}^{\ast 0})$ is set as $\Sigma^\ast(1381)^0~(\bar{\Sigma}^\ast(1381)^0)$.

From Fig.~\ref{fig:dalitz16701381}, it is evident that the triangle singularity effects occurring in this process will result in two intersecting bands in both the $\bar{\Lambda} \pi - \Lambda \pi$ and $\Lambda \bar{\Lambda} - \Lambda \pi$ Dalitz plots.
Therefore, as a result of this intersection, the peak in the $\bar{\Lambda} \pi$ invariant mass spectrum will be significantly enhanced, increasing the likelihood of observing this peak in experiments.
In the bottom-left subfigure, a minor cusp emerges around 1.33 GeV, coinciding with the location of the $\Sigma \pi$ threshold.
A sudden decrease around 1.82 GeV is also observed, attributed to the limitation of phase space.
%
%Although these two phenomena are hardly observable, they indicate that the peak is truly enhanced.

In the bottom-right panel of Fig.~\ref{fig:dalitz16701381}, when only Fig.~\ref{fig:mechanism}(b) is taken into account, a flat plateau will be observed in the $\Lambda\bar{\Lambda}$ invariant mass spectrum from 2.6 GeV to 2.9 GeV. 
Conversely, as a result of the intersection of the two bands, a small peak will emerge on the plateau, located around 2.85 GeV.

\begin{figure}[htbp]
  \centering
  \includegraphics[width=1.0\linewidth]{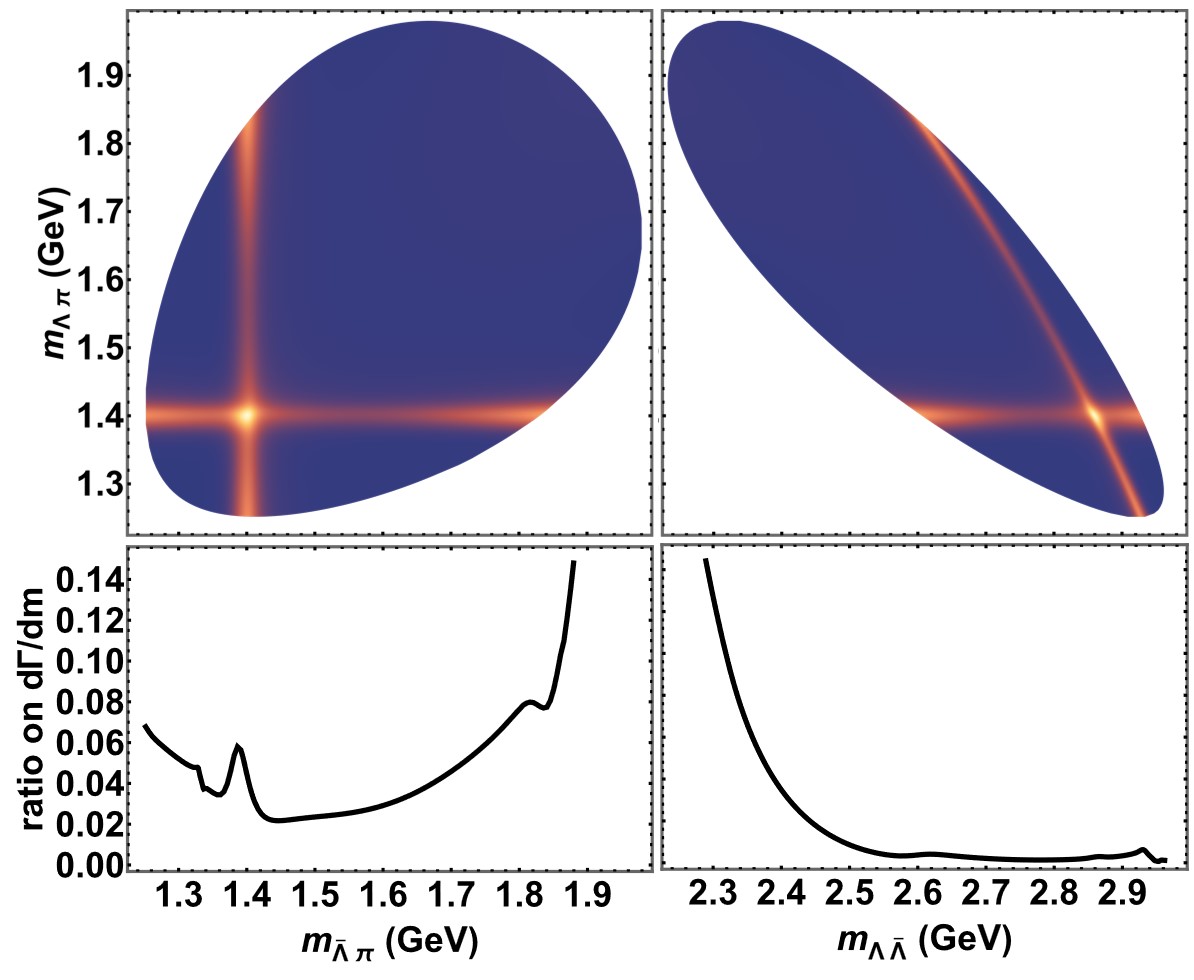}
\caption{Similar contents as given in Fig.~\ref{fig:dalitz16701381}, where $\Sigma^{\ast \pm}~(\bar{\Sigma}^{\ast \pm})$ in Fig.~\ref{fig:mechanism} is still $\Sigma^\ast(1670)^\pm~(\bar{\Sigma}^\ast(1670)^\pm)$ and the $\Sigma^{\ast 0}~(\bar{\Sigma}^{\ast 0})$ is changed from $\Sigma^\ast(1381)^0~(\bar{\Sigma}^\ast(1381)^0)$ to $\Sigma^\ast(1385)^0~(\bar{\Sigma}^\ast(1385)^0)$. 
In the above row, the figures are just Dalitz plots same as Fig.~\ref{fig:dalitz16701381}. 
While in the bottom row, "ratio on $d\Gamma/dm$" means that the figure draws the relative magnitude of the invariant mass spectra between $\Sigma^\ast(1670)^\pm \Sigma^{\ast}(1385)^0~(\bar{\Sigma}^\ast(1670)^\pm \bar{\Sigma}^{\ast}(1385)^0)$ case and $\Sigma^\ast(1670)^\pm \Sigma^{\ast}(1381)^0~(\bar{\Sigma}^\ast(1670)^\pm \bar{\Sigma}^{\ast}(1381)^0)$ case.}
\label{fig:dalitz16701385}
\end{figure}

Although Fig.~\ref{fig:dalitz16701381} has shown the basic phenomenon of our predicted triangle singularity, the currently missing status of $\Sigma^\ast(1381)$ indicates that choosing different $\Sigma^{\ast 0}~(\bar{\Sigma}^{\ast 0})$ to study the phenomena must be necessary.
%because practical experimental environments must be considered. 
%
Thus, in Fig.~\ref{fig:dalitz16701385}, $\Sigma^\ast(1381)$ is changed to the well established $\Sigma^\ast(1385)$ and similar contents as Fig.~\ref{fig:dalitz16701381} are presented.

As shown in Fig.~\ref{fig:dalitz16701385}, when $\Sigma^{\ast 0}~(\bar{\Sigma}^{\ast 0})$ changes from $\Sigma^\ast(1381)^0~(\bar{\Sigma}^\ast(1381)^0)$ to $\Sigma^\ast(1385)^0~(\bar{\Sigma}^\ast(1385)^0)$, the lineshapes change a lot. 
From the plots in the first row, the bands in the Dalitz plots are no longer uniform, indicating a typical $P-$wave contribution.
From the bottom row, it can also be observed that both the $\bar{\Lambda}\pi$ and $\Lambda \bar{\Lambda}$ invariant mass spectra are significantly suppressed. 
The changes can be attributed to the $J^P$ quantum number of the exchanged $\Sigma^{\ast 0}~(\bar{\Sigma}^{\ast 0})$.
If the quantum number of $\Sigma^\ast(1381)$ is $J^P=(\frac{1}{2})^-$, the interactions between $\Sigma^\ast(1381)$ and 
$\Sigma\pi$ as well as $\Lambda\pi$ are both in $S-$wave. 
For the $\Sigma^\ast(1385)$, the quantum number is $J^P=(\frac{3}{2})^+$, indicating that both its interactions with $\Sigma\pi$ and $\Lambda\pi$ transition to $P-$wave.
The interactions in higher partial waves may diminish the impact of triangle singularities and threshold cusps, a conclusion that has been previously explored and presented in Refs.~\cite{Swanson:2014tra, Shen:2020gpw}.

\begin{figure}[htbp]
  \centering
  \includegraphics[width=1.0\linewidth]{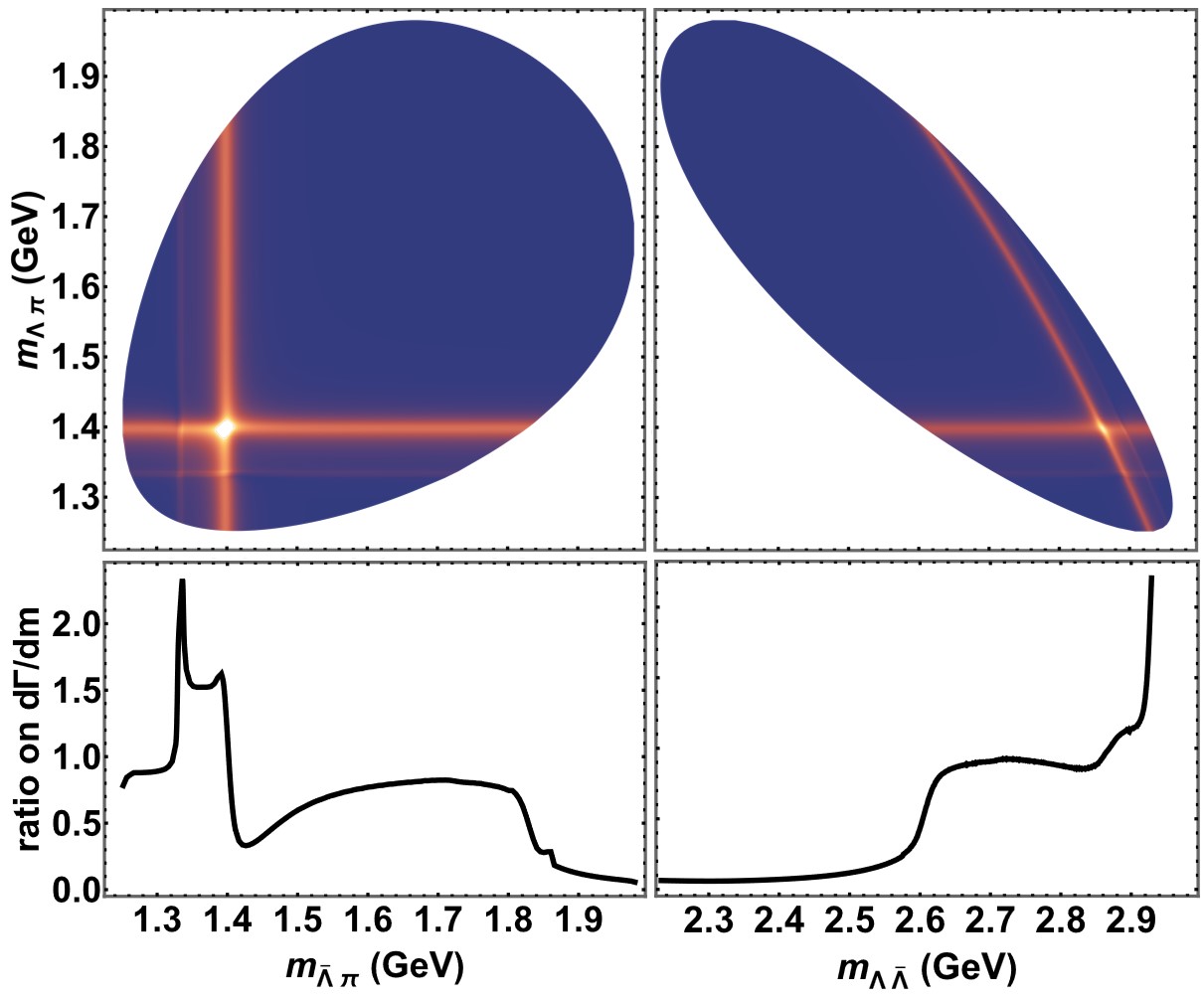}
\caption{Similar contents as given in Fig.~\ref{fig:dalitz16701381}, where $\Sigma^{\ast \pm}~(\bar{\Sigma}^{\ast \pm})$ in Fig.~\ref{fig:mechanism} is changed to $\Sigma^\ast(1620)^\pm~(\bar{\Sigma}^\ast(1620)^\pm)$ and the $\Sigma^{\ast 0}~(\bar{\Sigma}^{\ast 0})$ is still $\Sigma^\ast(1381)^0~(\bar{\Sigma}^\ast(1381)^0)$. 
In the above row, the figures are just Dalitz plots the same as Fig.~\ref{fig:dalitz16701381}. 
While in the bottom row, "ratio on $d\Gamma/dm$" means that the figure draws the relative magnitude of the invariant mass spectra between $\Sigma^\ast(1620)^\pm \Sigma^{\ast}(1381)^0~(\bar{\Sigma}^\ast(1620)^\pm \bar{\Sigma}^{\ast}(1381)^0)$ case and $\Sigma^\ast(1670)^\pm \Sigma^{\ast}(1381)^0~(\bar{\Sigma}^\ast(1670)^\pm \bar{\Sigma}^{\ast}(1381)^0)$ case.}
\label{fig:dalitz16201381}
\end{figure}

To validate this hypothesis, it would be intriguing to examine an alternative configuration, where $\Sigma^{\ast 0}~(\bar{\Sigma}^{\ast 0})$ corresponds to $\Sigma^\ast(1381)^0~(\bar{\Sigma}^\ast(1381)^0)$, and $\Sigma^{\ast \pm}~(\bar{\Sigma}^{\ast \pm})$ corresponds to $\Sigma^\ast(1620)^\pm~(\bar{\Sigma}^\ast(1620)^\pm)$.
At this point, given that the $J^P$ values of $\Sigma^\ast(1670)$ and $\Sigma^\ast(1620)$ are $(\frac{3}{2})^-$ and $(\frac{1}{2})^-$ respectively, the interactions between $\Sigma^{\ast \pm}~(\bar{\Sigma}^{\ast \pm})$ and $\Lambda\pi^\pm(\bar{\Lambda}\pi^\pm)$ transition from $D-$wave to $S-$wave. These modifications are reflected in the results depicted in Fig.~\ref{fig:dalitz16201381}.

Upon comparing the panels in Fig.~\ref{fig:dalitz16701381} and Fig.~\ref{fig:dalitz16201381}, it becomes evident that when the interactions between $\Sigma^{\ast \pm}~(\bar{\Sigma}^{\ast \pm})$ and $\Lambda\pi^\pm(\bar{\Lambda}\pi^\pm)$ transition from $D-$wave to $S-$wave, the peak height is significantly enhanced.
Furthermore, two distinct bands also emerge, precisely positioned at the thresholds of $\Sigma \pi$ and $\bar{\Sigma} \pi$ respectively.
This means that the cusp effect at this time is enhanced a lot. 
Furthermore, from the invariant mass spectra in the bottom row, it can be seen that when the partial wave of interaction is reduced, the triangle singularity effects are enhanced a lot, with a factor around 2.0.
Thus, we may also get the conclusion that higher partial wave will suppress the effect of triangle singularity in addition with the cusp effect.

\subsection{Analysis on the background and the significance of triangle singularity effect}

\begin{figure}[hhtbp]
  \centering
  \includegraphics[width=1.0\linewidth]{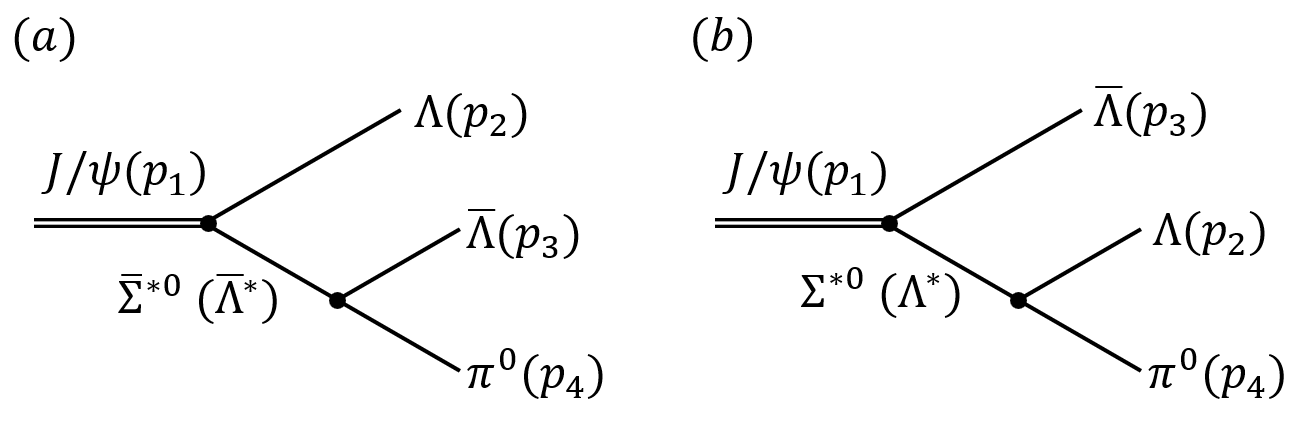}
\caption{The background of $J/\psi \to \Lambda \bar{\Lambda} \pi^0$ process considered in this work, where $(a)$ and $(b)$ are just conjugate to each other.}
\label{fig:background}
\end{figure}

If the triangle singularity effects mentioned in the previous subsection can be detected in experiments, discussions on the background are absolutely essential.
Similar to our previous works~\cite{Huang:2020kxf,Huang:2021olv}, the background contributions are depicted as shown in Fig.~\ref{fig:background}.

We illustrate the two possible mechanisms for this isospin process at the tree level in Fig.~\ref{fig:background}.
Clearly, in the background, either the interaction between $J/\psi$ and $\Lambda \bar{\Sigma}^{\ast 0}$ or the interaction between $\Lambda^\ast$ and $\Lambda\pi$ violates the isospin conservation law of strong interactions. 
%
%Therefore, these interactions must be of the electroweak type.
Therefore, these strength of interactions must be similar as that of the electroweak type.
According to Ref.~\cite{Donald:PA}, typically, the relative strengths of the strong, electromagnetic, and weak forces are approximately $1$, $10^{-2}$, and $10^{-7}$ respectively.
Hence, it is reasonable to assume that either the interaction between $J/\psi$ and $\Lambda \bar{\Sigma}^{\ast 0}$ or the interaction between $\Lambda^\ast$ and $\Lambda\pi$ occurs through electromagnetic process at the quark level.
Upon consulting RPP~\cite{ParticleDataGroup:2022pth}, we observe that the branching ratios for the $J/\psi \to \Sigma^{\ast}(1385)^- \bar{\Sigma}^+$ and $J/\psi \to \Sigma^{\ast}(1385)^0 \bar{\Lambda}$ processes are $(3.1 \pm 0.5) \times 10^{-4}$ and $<4.1 \times 10^{-6}$, respectively.
In other words, the ratio between them is at most 100.
Therefore, the inference regarding the type of interactions at the $J/\psi \Lambda \bar{\Sigma}^{\ast 0}$ and $\Lambda^\ast \Lambda \pi$ vertices is justified.

The aim of this study is to investigate whether the triangle singularity effect remains observable after accounting for the background. 
The key focus area is the invariant mass of $\Lambda\pi~(\bar{\Lambda}\pi)$ around 1.4 GeV.
Near this region, there are two well established resonances $\Lambda(1405)$ and $\Sigma^\ast(1385)$ for the diagrams shown in Fig.~\ref{fig:background}. 
However, apart from the $J/\psi \to \Sigma^{\ast}(1385)^0 \bar{\Lambda}$ process, the $J/\psi \to \Lambda(1405)\bar{\Lambda}$ and $\Lambda(1405) \to \Lambda \pi$ processes have not been observed~\cite{ParticleDataGroup:2022pth}. 
This implies that the background originating from $\Lambda(1405)$ cannot be calculated with high precision.
Fortunately, the branching ratio of $J/\psi \to \Lambda \bar{\Sigma}^- \pi^+$ have been measured by previous experiments as $(8.3 \pm 0.7) \times 10^{-4}$~\cite{ParticleDataGroup:2022pth}. 
If we assume that this process is dominated by $\Lambda(1405)$, the coupling constant $g_{J/\psi \Lambda(1405) \Lambda}$ then can be extracted. 
Taking into account the isospin-breaking effect of approximately $1\%$, we can calculate the contribution of $\Lambda(1405)$ to $J/\psi \to \Lambda \bar{\Lambda} \pi^0$.

Since the $J/\psi \to \Lambda \bar{\Sigma}^- \pi^+$ process primarily occurs through $\Lambda(1405)$, the triangle singularity effect remains prominent when factoring in the background contribution around 1.4 GeV in the $\Lambda\pi~(\bar{\Lambda}\pi)$ invariant mass spectrum. 
This suggests that detecting it in future experiments holds promise.

Regarding the loop diagram, experimental data on the branching ratios of the $J/\psi \to \Sigma^\ast(1670) \bar{\Sigma}$ and $J/\psi \to \Sigma^\ast(1620) \bar{\Sigma}$ processes are currently unavailable, and as a result, the corresponding coupling constants cannot be accurately estimated.
When comparing the $J/\psi \to \Sigma^\ast(1385) \bar{\Sigma}$, $J/\psi \to \Sigma^\ast(1670) \bar{\Sigma}$, and $J/\psi \to \Sigma^\ast(1620) \bar{\Sigma}$ processes, it is observed that the $J/\psi \to \Sigma^\ast(1385) \bar{\Sigma}$ process takes place in the $S-$wave, whereas the others occur in the $P-$wave.
The branching ratio of the $J/\psi \to \Sigma^\ast(1620/1670) \bar{\Sigma}$ process may be slightly lower than that of $J/\psi \to \Sigma^\ast(1385) \bar{\Sigma}$.
Taking into account that the phase space of the $J/\psi \to \Sigma^\ast(1385) \bar{\Sigma}$ process is slightly larger, this will also result in a slightly lower branching ratio for the $J/\psi \to \Sigma^\ast(1620/1670) \bar{\Sigma}$ processes.
Considering these two points, we assume that the branching ratio of the $J/\psi \to \Sigma^\ast(1385) \bar{\Sigma}$ process is approximately 5 times greater than that of the $J/\psi \to \Sigma^\ast(1620/1670) \bar{\Sigma}$ processes, i.e., $\frac{Br(J/\psi \to \Sigma^\ast(1620/1670) \bar{\Sigma})}{Br(J/\psi \to \Sigma^\ast(1385) \bar{\Sigma})} = \frac{1}{5}$.
As for the interactions between $\Sigma^\ast(1381)$ and $\Lambda\pi/\Sigma\pi$, we assume that they are similar to those of $\Sigma^\ast(1385)$ in order to extract the coupling constants.

All the coupling constants mentioned in this study can be compiled in Table~\ref{tab:ccknown} and Table~\ref{tab:ccunknown}. Table~\ref{tab:ccknown} consists of the coupling constants extracted from existing experimental data, while Table~\ref{tab:ccunknown} comprises the coupling constants derived from estimated branching ratios.

\begin{table}[htpb]
	\renewcommand\arraystretch{1.5}
  \centering
	\caption{The coupling constants extracted from RPP~\cite{ParticleDataGroup:2022pth}.}\label{tab:ccknown}
  \begin{tabular}{ccccc}
  \toprule[1pt]
  \iffalse
  Coupling constant &$~$& Branching Ratio &$~$& Value\\
  \fi
  Couplings &$~$& Branching Ratio &$~$& Value\\
  \midrule[1pt]
  \iffalse
  $g_{J/\psi\Sigma(1385)\Sigma}$ &$~$& $(3.1 \pm 0.5) \times 10^{-4}$ &$~$& $(2.738 \pm 0.221) \times 10^{-4}$ GeV$^{-1}$\\
  \fi
  $g_{J/\psi\Sigma(1385)\Sigma}$ &$~$& $(3.1 \pm 0.5) \times 10^{-4}$ &$~$& $(2.7 \pm 0.2) \times 10^{-4}$ GeV$^{-1}$\\
  $g_{J/\psi\Sigma(1385)\Lambda}$ &$~$& $<4.1 \times 10^{-6}$ &$~$& $<3.101 \times 10^{-5}$ GeV$^{-1}$\\
  $g_{\Sigma(1385)\Lambda\pi}$ &$~$& $(87.0 \pm 1.5)\%$ &$~$& $9.034 \pm 0.081$ GeV$^{-1}$\\
  $g_{\Sigma(1385)\Sigma\pi}$ &$~$& $(11.7 \pm 1.5)\%$ &$~$& $6.830 \pm 0.432$ GeV$^{-1}$\\
  $g_{\Sigma(1620)\Lambda\pi}$ &$\quad$& $(9.0 \pm 3.0)\%$ &$\quad$& $0.273 \pm 0.045$ \\
  $g_{\Sigma(1620)\Sigma\pi}$ &$\quad$& $(17 \pm 5)\%$ &$\quad$& $0.392 \pm 0.052$\\
  $g_{\Sigma(1670)\Lambda\pi}$ &$\quad$& $(10 \pm 5)\%$ &$\quad$& 3.713 $\pm$ 0.765 GeV$^{-2}$\\
  $g_{\Sigma(1670)\Sigma\pi}$ &$\quad$& $(45 \pm 15)\%$ &$\quad$& 10.299 $\pm$ 1.600 GeV$^{-2}$\\
  $g_{\Lambda(1405)\Sigma\pi}$ & $\quad$ & $\sim 100\%$ & $\quad$ & $\sim 1.569$\\
  $g_{J/\psi \Lambda(1405) \Lambda}$ & $\quad$ & $(8.3 \pm 0.7) \times 10^{-4}$ & $\quad$ & $(1.335 \pm 0.021) \times 10^{-3}$\\
  \bottomrule[1pt]
  \end{tabular}
\end{table}

\begin{table}[htpb]
	\renewcommand\arraystretch{1.5}
  \centering
	\caption{The estimated coupling constants calculated by the estimated branching ratios.}\label{tab:ccunknown}
  \begin{tabular}{ccccc}
  \toprule[1pt]
  Coupling constant &$\quad$& Branching Ratio &$\quad$& Value\\
  \midrule[1pt]
  $g_{\Lambda(1405)\Lambda\pi}$ &$\quad$& $\sim 1\%$ &$\quad$& $\sim 0.131$\\
  $g_{J/\psi \Sigma(1670) \Sigma}$ &$\quad$& $\sim 6.2 \times 10^{-5}$ &$\quad$& $\sim 7.129 \times 10^{-3}$\\
  $g_{J/\psi \Sigma(1620) \Sigma}$ &$\quad$& $\sim 6.2 \times 10^{-5}$ &$\quad$& $\sim 1.475 \times 10^{-3}$\\
  $g_{\Sigma(1381)\Lambda\pi}$ &$~$& $\sim 85 \%$ &$~$& $\sim 2.585$\\
  $g_{\Sigma(1381)\Sigma\pi}$ &$~$& $\sim 15 \%$ &$~$& $\sim 0.815$\\
  \bottomrule[1pt]
  \end{tabular}
\end{table}

\begin{figure}[htbp]
  \centering
  \includegraphics[width=1.0\linewidth]{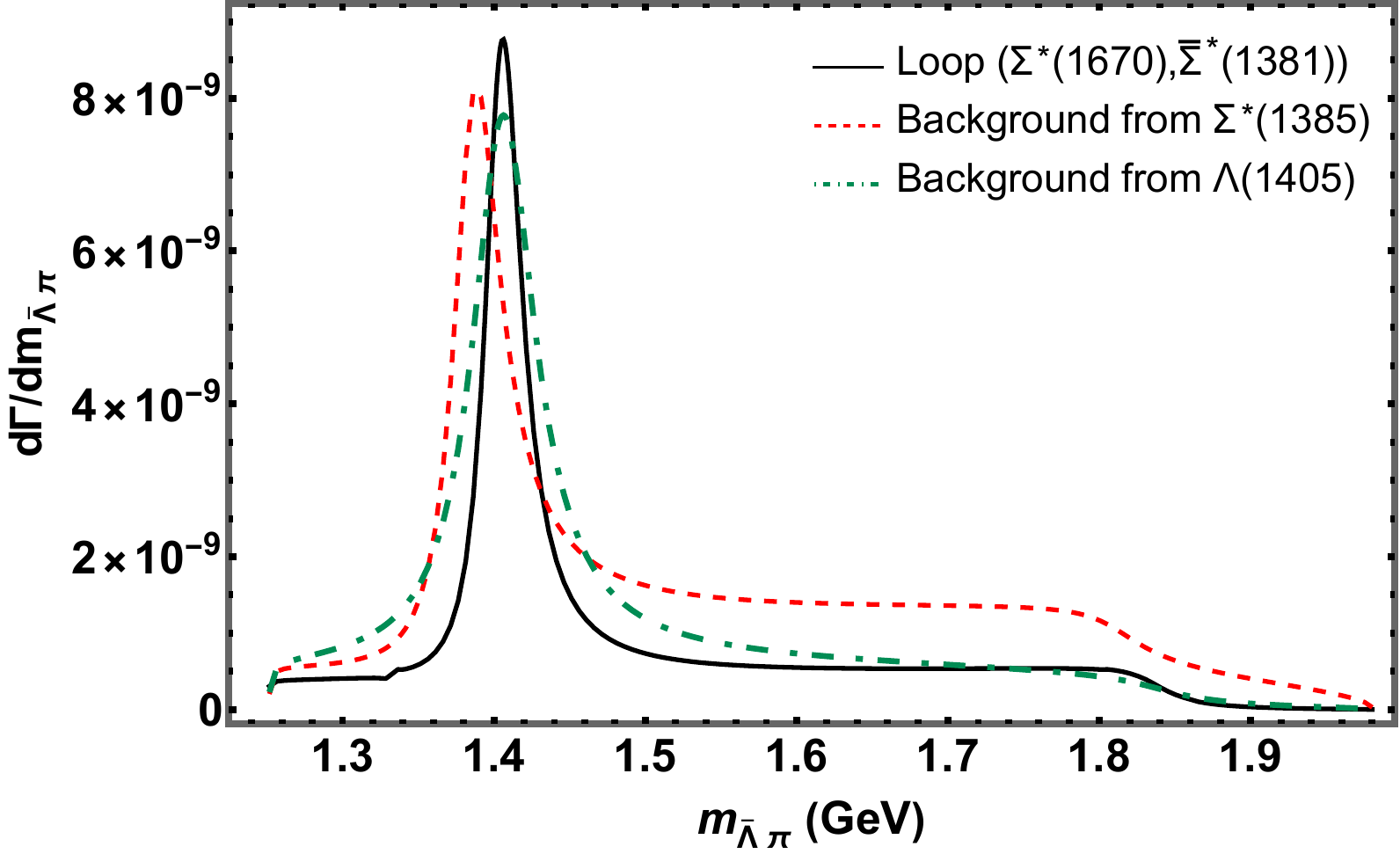}
\caption{The $\bar{\Lambda} \pi$ invariant mass spectrum of the background and loop diagram. Here, the black solid line is the contribution of loop diagram, where we set $\Sigma^{\ast\pm}~(\bar{\Sigma}^{\ast\pm})$ as $\Sigma^{\ast}(1670)^\pm~(\bar{\Sigma}^{\ast}(1670)^\pm)$ and $\Sigma^{\ast0}~(\bar{\Sigma}^{\ast0})$ as $\Sigma^{\ast}(1381)^0~(\bar{\Sigma}^{\ast}(1381)^0)$. The red and green dashed lines are the contributions of background, where the intermediate states are $\Sigma^\ast(1385)$ and $\Lambda(1405)$, respectively.}
\label{fig:backgroundims}
\end{figure}

With the these coupling constants, the significance of triangle singularity related to the background can be estimated. 
In Fig.~\ref{fig:backgroundims}, as an example, we display the $\bar{\Lambda} \pi$ invariant mass spectrum, which includes the contributions of both background and the loop diagram with the $\Sigma^\ast(1670) \Sigma^\ast(1381) ~(\bar{\Sigma}^\ast(1670) \bar{\Sigma}^\ast(1381))$ configuration.
In the background, we include both the contributions from $\Lambda(1405)$ and $\Sigma^\ast(1385)$ as intermediate states.

As shown in Fig.~\ref{fig:backgroundims}, around 1.4 GeV, the peak position of the triangle singularity is nearly identical, though slightly higher, than the background attributed to $\Lambda(1405)$. 
It is clear that the contribution from the background is suppressed, likely due to the background originating from an isospin-breaking process. 
However, the contribution from the loop diagram is almost the same as that of tree diagram. 
After integrating the $\bar{\Lambda}\pi$ invariant mass spectrum, we have determined the branching ratios for the loop, the $\Sigma(1385)$ and the $\Lambda(1405)$ background components to be approximately $6.76 \times 10^{-6}$, $1.18 \times 10^{-5}$, and $9.27 \times 10^{-6}$, respectively. 
This indicates that the contributions of the triangle singularity and background are comparable.
Although the position of the triangle singularity coincides with that of the $\Lambda(1405)$, it is evident from Fig.~\ref{fig:backgroundims} that their lineshapes differ. 
This results in the triangle singularity effect being much sharper than the resonance structure. 
This results in the triangle singularity effect being much sharper than the resonance structure.

After estimating the backgrounds, the significance of the triangle singularity is determined using the variable
\begin{eqnarray}
  \mathrm{ratio}(m_{\bar{\Lambda}\pi}) = \frac{(d\Gamma/dm_{\bar{\Lambda}\pi})_{\mathrm{TS}}}{(d\Gamma/dm_{\bar{\Lambda}\pi})_{\mathrm{background}}},\label{eq:ratio}
\end{eqnarray}
which represents the relative magnitude between the triangle singularity and background in the $\bar{\Lambda}\pi$ invariant mass spectrum.
This ratio excludes the interference contamination from the loop diagrams and background, and it circumvents the phase angle issue as discussed in Refs.~\cite{Huang:2020kxf,Huang:2021olv}, as mentioned in the introduction. 
If the triangle singularity effect can still be identified from this ratio, its detection in experiments is assured.

\begin{figure}[htbp]
  \centering
  \includegraphics[width=1.0\linewidth]{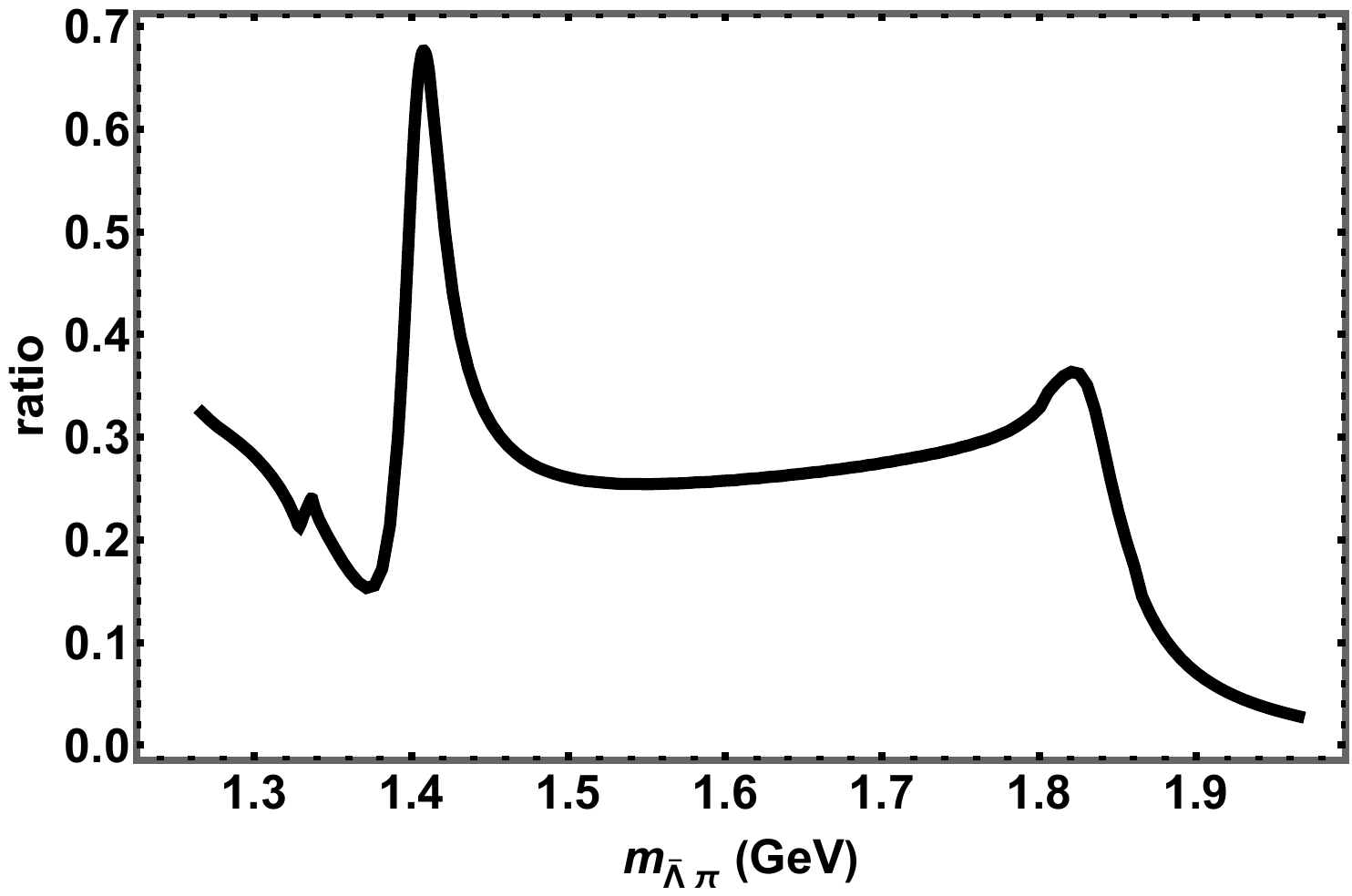}
\caption{The relative contribution between triangle singularity effect and background on $\bar{\Lambda}\pi$ invariant mass spectrum, where the $\Sigma^{\ast \pm}~(\bar{\Sigma}^{\ast \pm})$ and $\Sigma^{\ast 0}~(\bar{\Sigma}^{\ast 0})$ in Fig.~\ref{fig:mechanism} are set as $\Sigma^\ast(1670)^\pm~(\bar{\Sigma}^\ast(1670)^\pm)$ and $\Sigma^\ast(1381)^0~(\bar{\Sigma}^\ast(1381)^0)$ respectively.}
\label{fig:ratio16701381}
\end{figure}

\begin{figure}[htbp]
  \centering
  \includegraphics[width=1.0\linewidth]{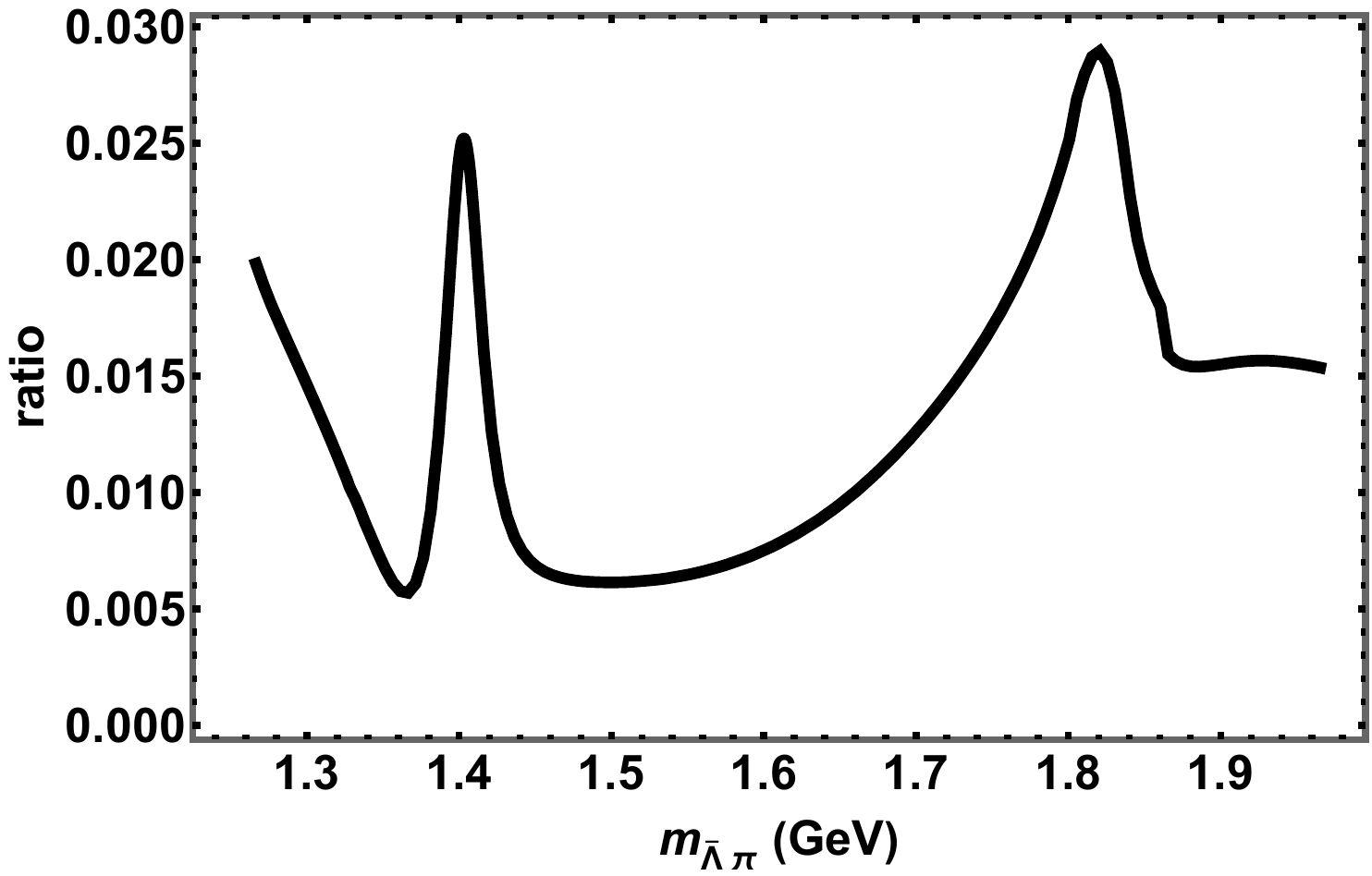}
\caption{The relative contribution between triangle singularity effect and background on $\bar{\Lambda}\pi$ invariant mass spectrum, where the $\Sigma^{\ast \pm}~(\bar{\Sigma}^{\ast \pm})$ and $\Sigma^{\ast 0}~(\bar{\Sigma}^{\ast 0})$ in Fig.~\ref{fig:mechanism} are set as $\Sigma^\ast(1670)^\pm~(\bar{\Sigma}^\ast(1670)^\pm)$ and $\Sigma^\ast(1385)^0~(\bar{\Sigma}^\ast(1385)^0)$ respectively.}
\label{fig:ratio16701385}
\end{figure}

\begin{figure}[htbp]
  \centering
  \includegraphics[width=1.0\linewidth]{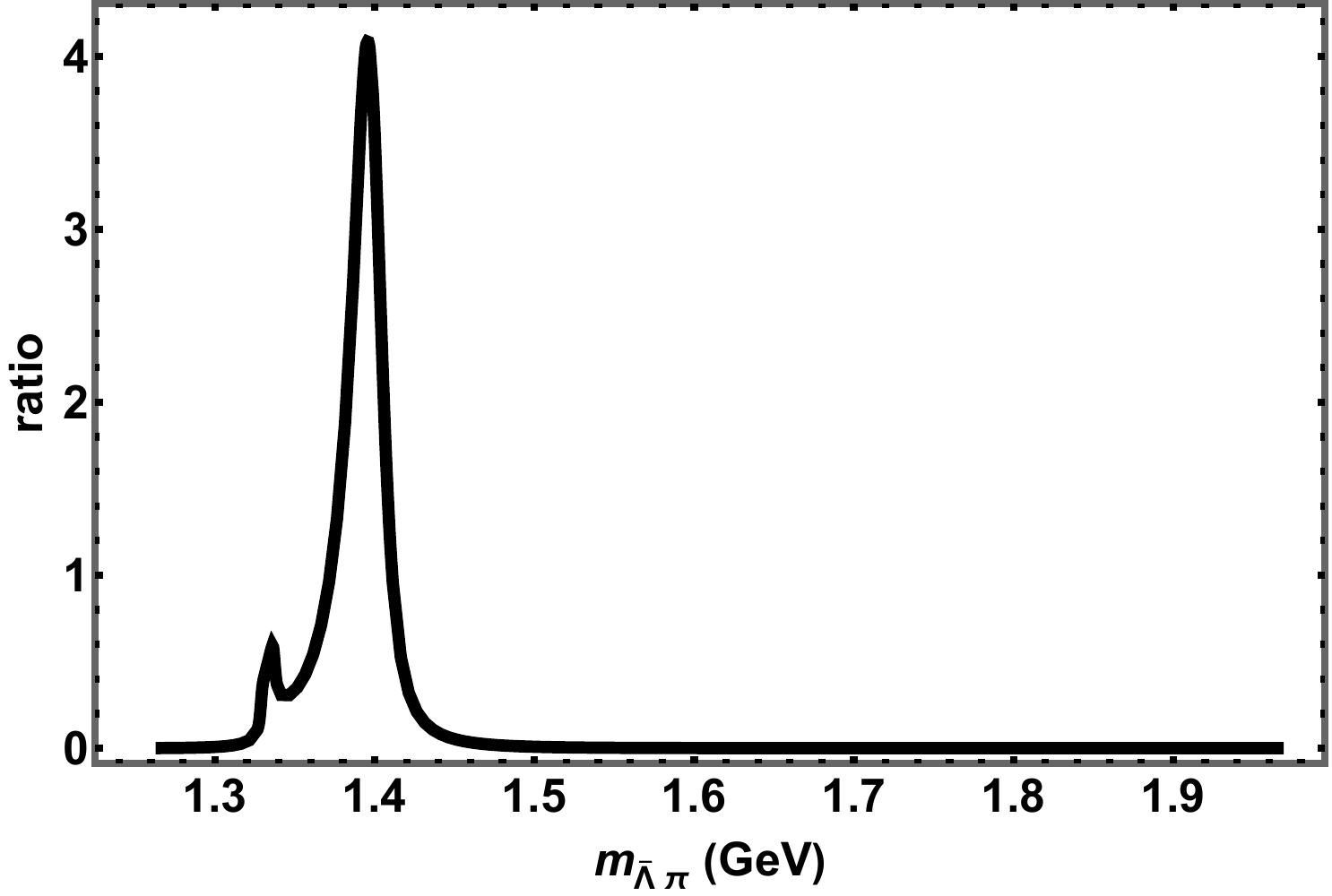}
\caption{The relative contribution between triangle singularity effect and background on $\bar{\Lambda}\pi$ invariant mass spectrum, where the $\Sigma^{\ast \pm}~(\bar{\Sigma}^{\ast \pm})$ and $\Sigma^{\ast 0}~(\bar{\Sigma}^{\ast 0})$ in Fig.~\ref{fig:mechanism} are set as $\Sigma^\ast(1620)^\pm~(\bar{\Sigma}^\ast(1620)^\pm)$ and $\Sigma^\ast(1381)^0~(\bar{\Sigma}^\ast(1381)^0)$ respectively.}
\label{fig:ratio16201381}
\end{figure}

Using the ratio determined from Eq.~(\ref{eq:ratio}), one can plot the distributions in $[\Sigma^{\ast\pm},\Sigma^{\ast0}]=[\Sigma^\ast(1670)^\pm,\Sigma^\ast(1381)^0]$, $[\Sigma^\ast(1670)^\pm,\Sigma^\ast(1385)^0]$, $[\Sigma^\ast(1620)^\pm,\Sigma^\ast(1381)^0]$ cases. The results are presented in Fig.~\ref{fig:ratio16701381}, Fig.~\ref{fig:ratio16701385}, and Fig.~\ref{fig:ratio16201381}, respectively.

Figs.~\ref{fig:ratio16701381}-\ref{fig:ratio16201381} display the relative contribution from the triangle singularity for all $[\Sigma^{\ast\pm},\Sigma^{\ast0}]$ cases. 
It is evident that a peak around 1.4 GeV emerges in the $\bar{\Lambda}\pi$ invariant mass spectrum, despite the significant differences in lineshapes.
The narrow peaks are a result of the widths of the triangle singularity and the peak in the background being 20 MeV and 60 MeV, respectively.
This indicate that the triangle singularity effect is significant compared to the background, 
and the signal peak of triangle singularity is much sharper than the resonance structure.
Especially, for the $\Sigma^\ast(1620)^\pm~(\bar{\Sigma}^\ast(1620)^\pm)$ and $\Sigma^\ast(1381)^0~(\bar{\Sigma}^\ast(1381)^0)$ cases, the contribution of the triangle singularity at the peak is approximately four times larger than that of the resonance, as illustrated in Fig.~\ref{fig:ratio16201381}.

\begin{figure}[htbp]
  \centering
  \includegraphics[width=1.0\linewidth]{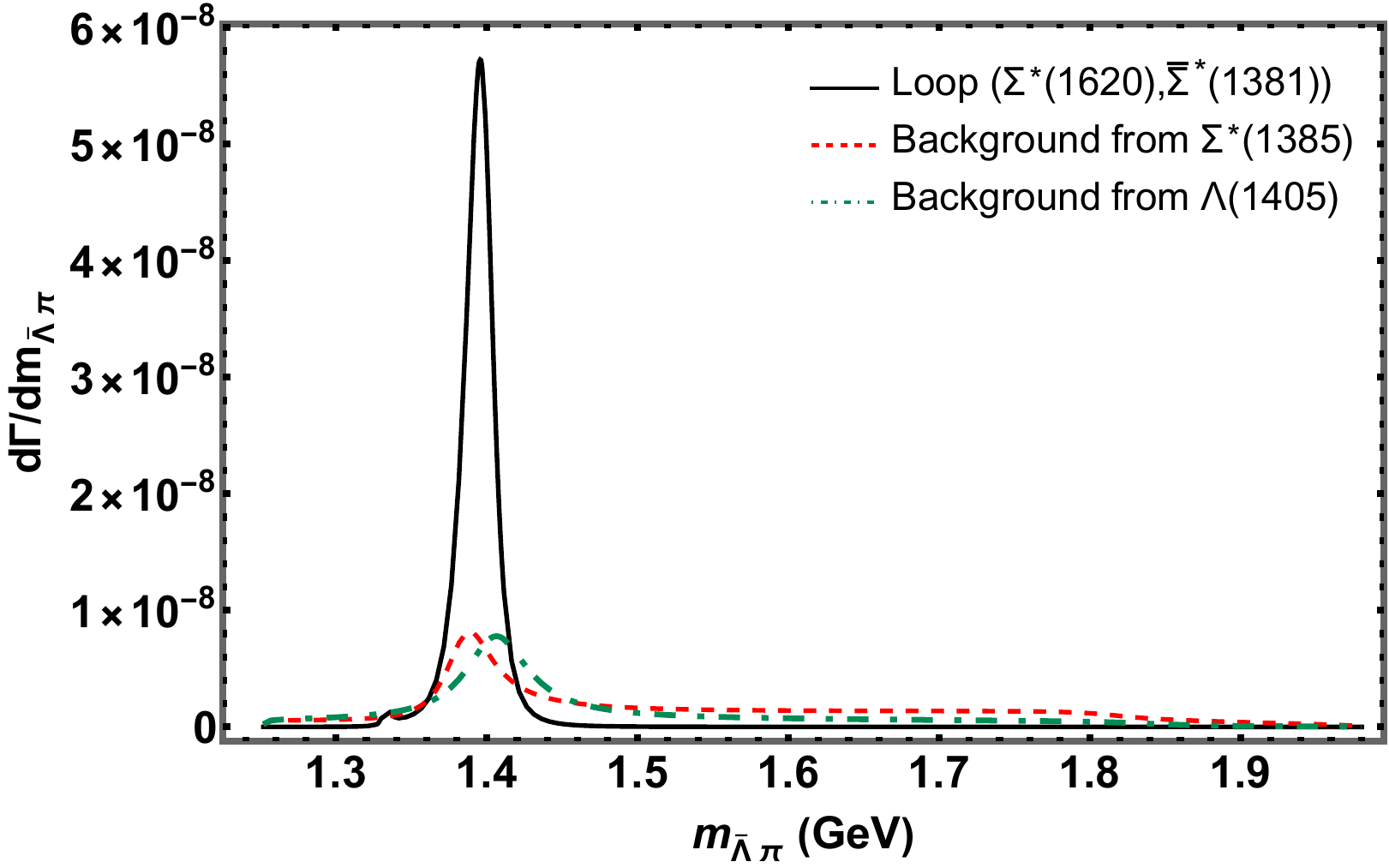}
\caption{The $\bar{\Lambda} \pi$ invariant mass spectrum of the background and loop diagram. Here, the black solid line means the contribution of background as in Fig.~\ref{fig:background}, where the intermediate states include $\Lambda(1405)$ and $\Sigma^\ast(1385)$. While for the contribution of loop diagram, we set $\Sigma^{\ast\pm}~(\bar{\Sigma}^{\ast\pm})$ as $\Sigma^{\ast}(1620)^\pm~(\bar{\Sigma}^{\ast}(1620)^\pm)$ and $\Sigma^{\ast0}~(\bar{\Sigma}^{\ast0})$ as $\Sigma^{\ast}(1381)^0~(\bar{\Sigma}^{\ast}(1381)^0)$.}
\label{fig:loop16201381}
\end{figure}

In Fig.~\ref{fig:loop16201381}, we illustrate the peak structure from the loop and background, with the triangle singularity peak dominating in the region around 1.4 GeV.
The ratio of each background is much smaller than that of the loop diagrams, and the branching ratios of the loop diagrams and all the backgrounds to $J/\psi$ are estimated to be approximately $1.56 \times 10^{-5}$ and $2.4 \times 10^{-5}$, respectively.
It has been observed that there is a significant contribution from the long tails to the tree diagrams.
This phenomenon suggests that if both $\Sigma^\ast(1620)$ and $\Sigma^\ast(1381)$ indeed exist, the peak around 1.4 GeV in the $\bar{\Lambda}\pi$ invariant mass spectrum may be predominantly influenced by the triangle singularity effect.

\subsection{The possible existence of missing $\Sigma^\ast(\frac{1}{2}^-)$}

In above calculations, we have included two $\Sigma^\ast(\frac{1}{2}^-)$ states in the singularity calculation.
One is the not-well-established $\Sigma^\ast(1620)$, the other is the still missing $\Sigma^\ast(1381)$. 
From Figs.~\ref{fig:ratio16701381}-\ref{fig:ratio16201381}, it is evident that both of their effects on the final results are substantial. 
For $\Sigma^\ast(1381)$, the triangle singularity effect attributed to it is more significant than that caused by $\Sigma^\ast(1385)$. 
This can be observed by comparing the lineshapes in Fig.~\ref{fig:ratio16701381} and Fig.~\ref{fig:ratio16701385}.
This is the outcome of the interactions of $\Sigma^\ast(1381)$ and $\Sigma^\ast(1385)$ with $\Sigma\pi/\Lambda\pi$ transitioning from $S-$wave to $P-$wave, respectively. 
Consequently, the triangle singularity effect will be greatly suppressed for the $P-$wave case, indicating that the influence of the triangle singularity generated via $\Sigma^\ast(1385)$ has almost no impact on the width of the entire peak.

Regarding $\Sigma^\ast(1620)$, its presence directly results in the triangle singularity when $\Sigma^{\ast0}$ is considered to be the $\Sigma^\ast(1381)$. 
The $\Sigma^\ast(1670) \Lambda\pi$ interaction occurs in $D-$wave, while that of $\Sigma^\ast(1620) \Lambda\pi$ takes place in $S$-wave. 
If one replaces $\Sigma^\ast(1670)$ with $\Sigma^\ast(1620)$, it will significantly enhance the triangle singularity. 
Therefore, the loop diagram will have the most significant impact on the $J/\psi \to \Lambda \bar{\Lambda} \pi^0$ process around 1.4 GeV in the $\Lambda\pi~(\bar{\Lambda}\pi)$ invariant mass spectrum.

In conclusion, if future experiments can confirm the predictions of this study, it will not only validate the concept of triangle singularity but also provide a secondary investigation into the existence of the two $\Sigma^\ast(\frac{1}{2}^-)$ states around 1.4 GeV and 1.6 GeV.

\section{Summary}\label{sec4}

While the concept of triangle singularity was introduced in 1959~\cite{Landau:1959fi}, a definitive investigation on it is still lacking to this day. 
To address this issue, our previous studies have suggested certain processes in which pure triangle singularities could potentially be observed in future experiments~\cite{Huang:2020kxf,Huang:2021olv}.

In our previous studies, we observed that the strength of the triangle singularity is influenced by its interference with the background, which is attributed to the limited impact of the triangle singularity.
However, the interference phase angle is completely undetermined, which could hinder the observation of the triangle singularity effect in the experiment.
%
%In addition, the demands on the high luminosity and resolution of detector will make the observation and analysis difficult. 
%
To address these issues, we suggest a new process $J/\psi \to \Lambda \bar{\Lambda} \pi^0$, where the mass discrepancy between $\Sigma^{(\ast)}$ particles of varying charges will induce a triangle singularity with a width of approximately 20 MeV. 
This approach effectively resolves the resolution problem.
In particular, this process involves isospin breaking, resulting in a significant suppression of the background. 
As illustrated in Figs. \ref{fig:ratio16701381}-\ref{fig:ratio16201381}, the triangle singularity can disregard the phase angle. 
This implies that the observation of the triangle singularity effect can be more pronounced in practice.
Given the substantial number of $J/\psi$ events collected at BESIII and the anticipated future STCF, the predictions made in this study are more likely to be confirmed.

In addition to confirming the presence of the triangle singularity, the findings of this study also suggest the existence of a $\Sigma^\ast(\frac{1}{2}^-)$ state near 1.4 GeV, specifically the $\Sigma^\ast(1381)$ as previously predicted in Refs.~\cite{Zhang:2004xt,Wu:2009tu, Wu:2009nw}.
When compared to $\Sigma^\ast(1385)$, the interaction between $\Sigma^\ast(1381)$ and $\Lambda\pi/\Sigma\pi$ takes place in the $S-$wave, greatly amplifying the significance of the triangle singularity, as illustrated in Figs. \ref{fig:ratio16701381}-\ref{fig:ratio16201381}.
%
%For the not-well-established $\Sigma^\ast(1620)$, its contribution is also important, which may assure the existence of triangle singularity when cooperating with $\Sigma^\ast(1381)$.
For the not-well-established $\Sigma^\ast(1620)$, replacing $\Sigma^\ast(1670)$ with this particle will make the triangle singularity effect more pronounced, especially cooperating with $\Sigma^\ast(1381)$.

Thus, we encourage future experiments perform a detailed analysis on the $\Lambda\pi~(\bar{\Lambda}\pi)$ invariant mass spectrum of $J/\psi \to \Lambda \bar{\Lambda} \pi^0$ process, especially around 1.4 GeV. 
We recommend that experimentalist and theorists conduct further research on the $\Sigma^\ast(\frac{1}{2}^-)$ spectrum, as this will greatly aid in the verification of triangle singularities.

\section{Acknowlegment}

Qi Huang wants to thank Jia-lun Ping and Rui Chen for very useful discussions. Qi Huang also wants to thank Wei Zhuang for useful technical support.
This work is supported by the National Natural Science Foundation of China under Grant Nos. 12175239, 12221005, 12175244, and 11675080,
and by National Key Research and Development Program of China under Contracts 2020YFA0406400.
Q. H  is also supported by the Start-up Funds of Nanjing Normal University under Grant No. 184080H201B20.
J. J. W is also supported by Chinese Academy of Sciences under Grant No. YSBR-101, 
and by Xiaomi Foundation / Xiaomi Young Talents Program.

\vfil

\end{document}